# An innovative approach to structural instabilities in tetragonal tungsten bronze crystals through the concept of rigid unit modes

M. Smirnov[1]* and P. Saint-Grégoire[2]

[1] Saint-Petersburg State University, Physical Department, Petrodvoretz, 194508 St.-Petersburg, Russia

[2] University of Nîmes, Laboratoire MIPA, 30021 NIMES cédex, France

* E-mail : smirnomb@rambler.ru

**Abstract.** Tetragonal tungsten bronze (TTB) oxides are one of the most important classes of ferroelectrics. Many of these framework structures undergo ferroelastic transformations related to octahedron tilting deformations. Such tilting deformations are closely related to the Rigid Unit Modes (RUMs). This paper discusses the whole set of RUMs in an ideal TTB lattice and possible crystal structures which can emerge owing to the condensation of some of them. Analysis of available experimental data for the TTB-like niobates lends credence to the obtained theoretical predictions.

## 1. Introduction

The theory of second-order structural phase transitions (SPT) is based on the idea of order parameter. In the case of displacive SPT, the order parameter is a vector defining cooperative atomic displacements which describe the lattice deformation during the transformation from one phase to another. This deformation coincides (or almost coincides) with the eigenvector of the soft mode and possesses minimal mechanical stiffness. Hence, in order to determine the atomistic pattern of the order parameter for a particular SPT one should reveal the softest degree of freedom of the crystal lattice.

The crystal structures built on oxide frameworks are lattices formed by quasi-rigid $XO_n$ polyhedra (tetrahedra, octahedra etc.), interconnected by common vertices or edges. It is possible that in such a structure there exist collective atomic displacements related to polyhedron translations and rotations and not involving any polyhedron deformations. Such deformations (called Rigid Unit Modes, RUM) are not accompanied by length variations of the X-O bonds and the O-O contacts. Thus, they are mechanically soft and potentially could play the role of soft mode eigenvectors.

A classic example of a SPT in which the soft mode is a RUM is the ferroelastic SPT in pervskite $AXO_3$ structure. The cubic perovskite (CP) structure is a framework of the corner-sharing regular octahedra. In this structure, all RUMs involve concerted rotations of the octahedra located within a layer perpendicular to the rotational axis which may be parallel to the *a*, *b* or *c* axes. The first example of the RUM-induced instability was the ferroelastic phase transition in $SrTiO_3$ [1]. Much attention was paid to the RUM-induced "compressibility collapse" in $ReO_3$ [2]. By now, the family of the perovskite-like ferroelastics enlarged considerably. It includes various complex oxides (titanates, zirconates, aluminates, tantalates etc) which display a wide variety of physical properties such as superconductivity, magnetism, ferroelectricity, and magnetoelectricity which make them very interesting for applications [3].

Tetragonal tungsten bronze oxides constitute one of the most important classes of ferroelectrics next to perovskites. They also can be viewed as framework lattices consisting of corner-sharing octahedra $NbO_6$. However, the manner of the octahedra arrangement within a layer of the TTB structure (see Fig. 1a) differs from that in a perovskite structure, first by its lower symmetry. In $c$-direction, the layers within a TTB lattice are repeatedly interconnected by sharing the apex octahedron corners. It is seen that within this layer-like framework lattice there are pentagonal, square, and trigonal tunnels which can accommodate cations of different sizes – respectively denoted as $M^{(5)}$, $M^{(4)}$ and $M^{(3)}$. Thus the net formula can be written as $M_i^{(5)} M_j^{(4)} M_k^{(3)} NbO_3$. Due to the large choice of the inserted cations, the TTB family presents a rich sequence of phase transitions and includes a large number of functional crystals and materials possessing electrooptic, pyroelectric and piezoelectric properties [5].

Many TTB structures undergo ferroelastic SPTs which transform the tetragonal lattice into orthorhombic ones. The tetragonal-to-orthorhombic distortions are usually weak and in some cases manifest themselves via incommensurate (INC) structure modulations. It seems reasonable to suggest that the ferroelastic transformations in the TTB crystals can also be related to the RUM tilting deformations. In this crystalline family, the important role of octahedron tilting was emphasized [6]. However, a detailed analysis of the whole RUM spectrum in the TTB structure and the variety of possible crystal structures induced by their condensation was not yet done.

This paper aims at filling this gap. It is organized as follows. First, we describe the whole set of RUMs in an ideal TTB lattice. Then, the crystal structures resulting from the starting tetragonal TTB-lattice owing to condensation of different RUMs are specified. Finally, these results are confronted with available experimental information for the niobate TTB-like crystals.

## 2. Method

The whole set of the RUM-induced structures originated from cubic perovskite was described by Glazer [4]. As an elementary RUM deformation he considered the concerted octahedron tilting around one of lattice axis. The octahedron rigidity condition strictly connects the tilting of octahedra localized within the same layer perpendicular to the axis of rotation but does not constrain rotations of octahedra localized within different such layers. Thus the octahedra of neighboring layers may tilt in-phase, anti-phase or even with arbitrary phase shift. Glazer introduced notations a+, a-, b+, b-, c+ and c- for the in-phase and anti-phase octahedron rotations around axes oriented along $a$, $b$ and $c$ axes. It is noteworthy, that the tilting deformations around mutually perpendicular axes are independent. Thus, any possible RUMs in a perovskite lattice can be determined by a triad $\phi_a, \phi_b, \phi_c$ with components determining the phase shifts between tilt angles of neighboring layers perpendicular to the $a$, $b$ and $c$ axes.

Depending on the combination of $\phi_a, \phi_b, \phi_c$ values, various crystal structures can be obtained from the cubic perovskite structure. A detailed analysis of such structures for the case $\phi_i = 0, \pi$ was presented in Ref. [4]. Many of them were really observed for particular perovskite-like compounds. In some cases, the octahedron tilting deformations really play the role of the order-parameter for ferroelastic phase transitions.

Later, the concept of concerted tilting of rigid polyhedra was generalized to be applicable for arbitrary framework structure [7]. In that paper, the notion of the RUM as a phonon mode which does not involve any structural distortions except polyhedron rotations and translations was clearly formulated. The question has arisen – how to determine the whole set of RUMs for an arbitrary crystal structure? The split-atom model was proposed to solve this problem. The essence of this approach is to view each polyhedron as a separate rigid body and to treat the linked corners as kept together by harmonic spring. We consider this method somewhat unnecessarily complicated, especially for the polyatomic structures. Instead, we propose to simulate the phonon states of a framework lattice built of the corner-linked $XO_n$ polyhedra within simple valence force model which assumes non-zero force constants only for the intra-polyhedron X-O bonds and the O-O polyhedron edges. The cations (other than X atoms) should be omitted. Phonon modes simulated within such model can be easily divided in two parts: ones of high frequency which involve polyhedron deformations and others of zero frequency that are RUMs. So, search for the whole RUM spectrum consists in scanning the phonon states throughout the entire Brillouin zone (BZ) and distinguishing the zero-frequency modes. This routine was realized by using LADY program [8].

## 3. Results and discussion
### 3.1. Atomistic pattern of RUMs in TTB

The starting aristotype structure is tetragonal with P4/mbm symmetry. Most of the TTB-like compounds crystallize in this structure at high temperature. The unit cell of this lattice and the corresponding BZ are shown in Fig 1a and 1b.

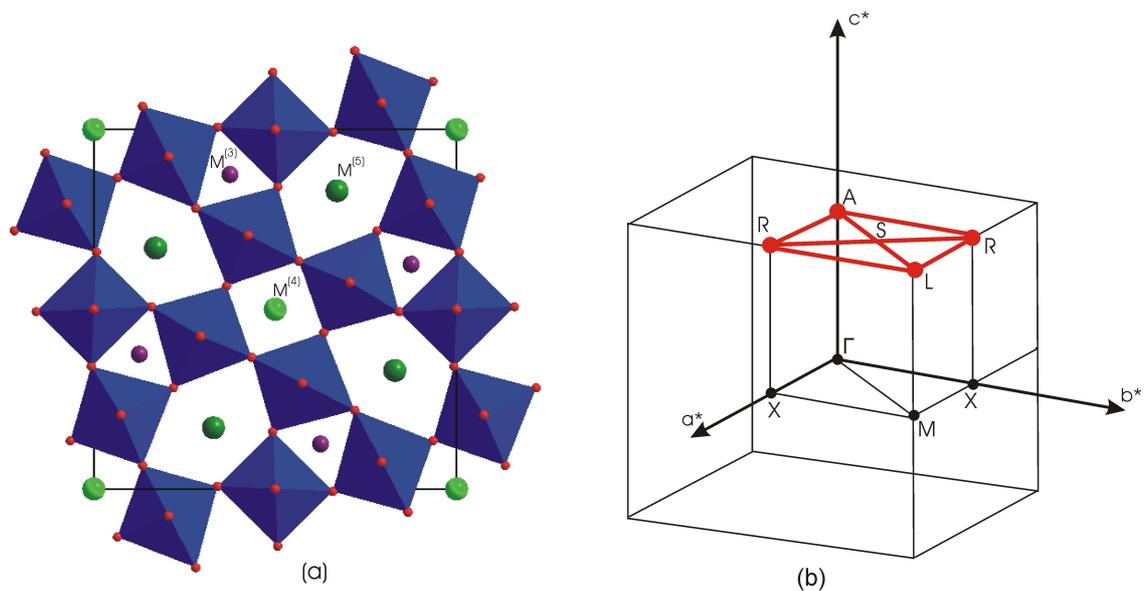

Figure 1. Structure (a) and BZ (b) of an ideal TTB crystal.

Special points within BZ are $Z = (0\,0\,\tfrac{1}{2})$, $A = (\tfrac{1}{2}\,\tfrac{1}{2}\,\tfrac{1}{2})$, $R = (0\,\tfrac{1}{2}\,\tfrac{1}{2})$ and $S = (\tfrac{1}{4}\,\tfrac{1}{4}\,\tfrac{1}{2})$. For all these points, our calculations revealed the existence of two RUMs. Moreover, the calculations indicated the presence of two RUM branches along the directions Z–A and R-S-R. Eigenvectors of the RUMs belonging to the special BZ points are shown in Fig. 2.

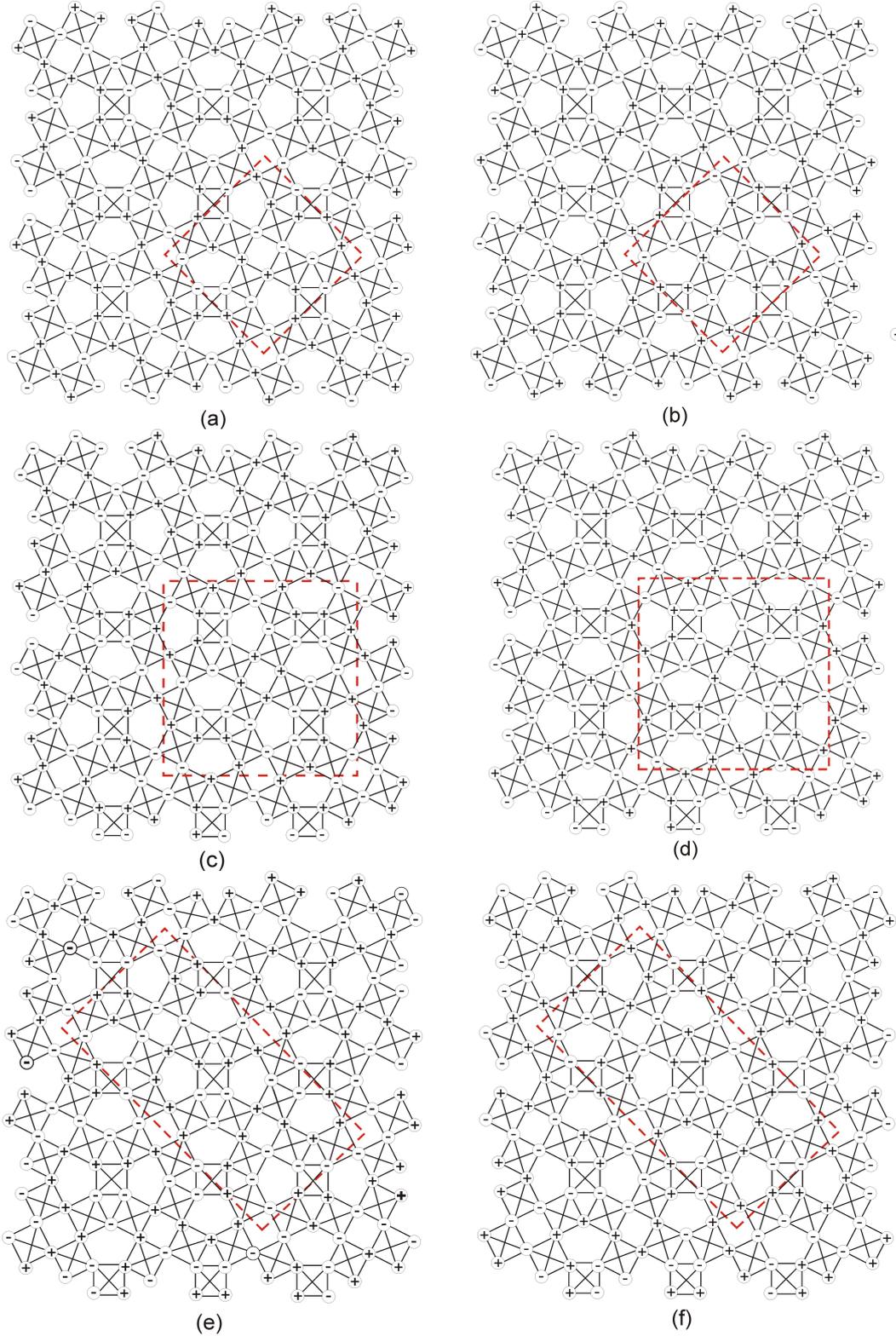

Figure 2. Eigenvectors of two Z-RUMs (a-b), A-RUMs (c-d) and R-RUMs (e-f). Dashed lines show unit cells of resulting substructures. The cations are not shown for simplicity.

Spatial patterns of these modes seem rather complicated. Nevertheless they can be classified in a simple way as shown below. A detailed analysis of the Z and A-RUMs shown in Figs. 2a-2d leads to conclusion that they can be represented as combinations of ρ-deformations shown in Fig. 3a. Analogously, the R-RUMs shown in Figs 2e-2f can be represented as combinations of σ-deformations shown in Fig. 3b.

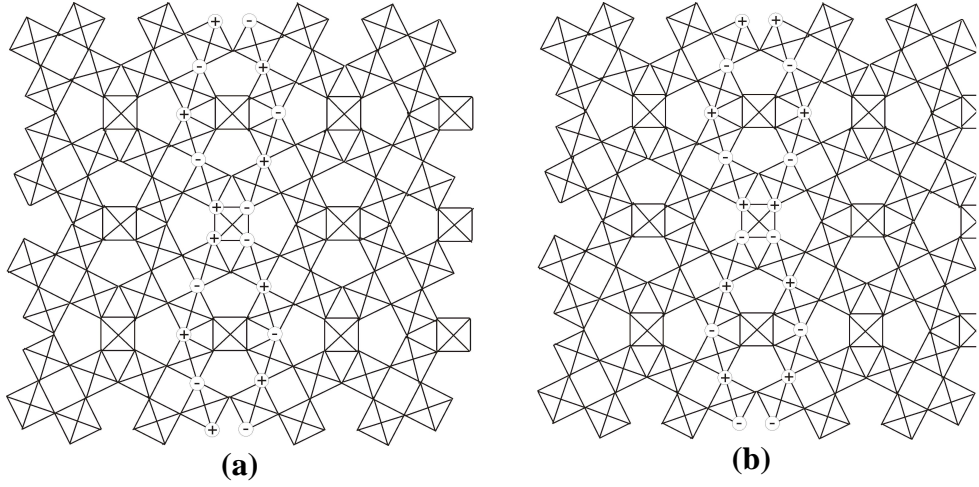

(a)  (b)

Figure 3. Two types of RUMs localized in the (110) layer: ρ-RUM (a) and σ-RUM (b).

It is remarkable that both ρ and σ-deformations shown in Fig. 3 are localized in layers perpendicular to (110) or $(1\bar{1}0)$ direction. Thus, arbitrary combinations of such ρ and σ-deformations localized in different layers would give rise to various RUMs. The situation is analogous to that in the CP lattice. The only difference is that in a CP lattice, due to the higher symmetry, the RUM containing layers are stacked in three directions (100), (010) and (001), whereas in TTB lattice they are can be staked only in two directions. It is important to note that the ρ deformations localized in mutually perpendicular layers do not overlap (in the sense that they involve displacements of different atoms). Hence, any combination of ρ deformations localized in (110) layers may be accompanied by arbitrary combination of the same deformations localized in the $(1\bar{1}0)$ layers thus giving rise to various RUMs. The same is valid for the σ deformations

Analyzing the CP lattice, Glazer considered the RUMs resulting from the in-phase and anti-phase combinations of octahedron rotations localized in neighboring layers. They were labeled by signs + and – attached to the vector determining the layer orientation. Similarly, one can use the notation $\sigma^+$, $\rho^+$, $\sigma^-$ and $\rho^-$ when describing the RUM deformations in a TTB lattice. These symbols determine the type and consequence of the concerted octahedron rotations. In order to specify completely the RUM distortion in a TTB lattice one must define these symbols for the two perpendicular directions (110) and $(1\bar{1}0)$. Below we use pairs of the symbols separated by a slash. Zero denotes the absence of deformations. Thus, the modes shown in Figs. 2a and 2b must be denoted as $\rho_+/\rho_+$ and $\rho_+/-\rho_+$; the modes in Figs 2c and 2d are denoted as $\rho^-/\rho^-$ and $\rho^-/-\rho^-$, respectively; and the modes in Figs. 2e and 2f as $\sigma^+/\sigma^+$ and $-\sigma^+/\sigma^+$.

These notations allow us to describe the RUM phonon branch along Z-A direction. This phonon branch in Z and A points corresponds to the $\rho_+/0$ and $\rho_-/0$ modes respectively. In the intermediate k-points the phonon of this branch corresponds to the ρ(φ)/0 modes where φ is the phase shift varying continuously between 0 and π. Similarly, the RUM branch along R-S-R direction which relates modes $0/\sigma_+$ and $0/\sigma_-$ consist of $0/\sigma(\phi)$ modes. The presence of the RUM phonon branches may result in an INC structures condensation with the modulation vectors $(\xi,\xi,0)$ and $(\xi,\overline{\xi},0)$.

*3.2. Substructures induced by RUM condensation*

In this section we analyze crystal structures resulting from the condensation of different RUMs (we call them *substructures*). RUMs related to different combinations of the ρ and σ rotations are listed in Tab. 1.

Table 1. Substructures resulting from the single-RUM condensation. The a, b, c vectors are unit cell vectors of the parent tetragonal structure and are unit cell vectors of resulting substructure

| $k$-point | RUM | $\mathbf{a}_s, \mathbf{b}_s, \mathbf{c}_s$ | Substructure |
|---|---|---|---|
| $Z = (0\,0\,\tfrac{1}{2})$ | $\rho_+/\rho_+$ | $\mathbf{a}, \mathbf{b}, 2\mathbf{c}$ | *Pbnm* (No. 62) |
| | $\rho_+/0$ | $\mathbf{a}-\mathbf{b}, \mathbf{a}+\mathbf{b}, 2\mathbf{c}$ | *Ccmm* (No. 63) |
| $A = (\tfrac{1}{2}\,\tfrac{1}{2}\,\tfrac{1}{2})$ | $\rho_-/\rho_-$ | $\mathbf{a}-\mathbf{b}, \mathbf{a}+\mathbf{b}, 2\mathbf{c}$ | *I*4/*m* (No. 87) |
| | $\rho_-/0$ | $\mathbf{a}-\mathbf{b}, \mathbf{a}+\mathbf{b}, 2\mathbf{c}$ | *Ibmm* (No. 74) |
| $R = (0\,\tfrac{1}{2}\,\tfrac{1}{2})$ | $\sigma_+/\sigma_+$ | $\mathbf{a}, 2\mathbf{b}, 2\mathbf{c}$ | $A2_1am$ (No. 36) |
| | $\sigma_+/0$ | $\mathbf{a}, 2\mathbf{b}, 2\mathbf{c}$ | $A112/m$ (No. 12) |

The structures resulting from the condensation of these RUMs are specified by defining their unit cell vectors and the space symmetry groups. It is interesting to consider substructures which result from simultaneous condensations of two RUM modes. Some of them are listed in Table 2.

Table 2. Substructures resulting from condensation of two RUMs

| $k$-points | RUM | $\mathbf{a}_s, \mathbf{b}_s, \mathbf{c}_s$ | Substructure |
|---|---|---|---|
| Z, A | $\rho_+/\rho_-$ | $\mathbf{a}-\mathbf{b}, \mathbf{a}+\mathbf{b}, 2\mathbf{c}$ | *Pcmn* (No. 62) |
| Z, R | $\rho_+/\sigma_+$ | $\mathbf{a}, 2\mathbf{b}, 2\mathbf{c}$ | $P112_1/m$ (No. 11) |
| A, R | $\rho_-/\sigma_+$ | $2\mathbf{a}, 2\mathbf{b}, 2\mathbf{c}$ | $P112/m$ (No. 10) |
| K, R | $\rho_+\rho_-/\sigma_+$ | $2(\mathbf{a}-\mathbf{b}), \mathbf{a}+\mathbf{b}, 2\mathbf{c}$ | *Bbmm* (No. 63) |
| K, K | $\rho_+\rho_-/\sigma_+\sigma_-$ | $2(\mathbf{a}-\mathbf{b}), 2(\mathbf{a}+\mathbf{b}), 2\mathbf{c}$ | $B112/m$ (No. 12) |

The structures listed in Tables 1-2 can be considered as resulting from the ideal tetragonal TTB lattice owing to the condensation of one or two RUMs, i.e. in consequence of a ferroelastic SPT. To our knowledge, no compound with the TTB structure undergoes a ferroelastic SPT directly from the tetragonal *P*4/*mbm* para-phase structure. For all such compounds, the ferroelastic SPT is preceded by a ferroelectric SPT. The sequence of SPTs observed on lowering temperature is usually as follows:

Para-phase → Ferroelectric → ferroelectric/ferroelastic

The low-temperature structures reported in literature correspond to ferroelectric-ferroelastic phases. The hypothetic substructures listed in Tables 1-2 correspond to structures resulting from para-phase via a ferroelastic SPT. In order to compare them with the structures determined experimentally, we must take into account structural distortions induced by ferroelectric SPTs. In fact, there is a good reason to think that structural distortions induced by ferroelastic and ferroelectric transformations are not coupled. The former consists indeed of concerted octahedron rotations (i.e. primarily involves displacements of oxygen atoms), and the latter involves primarily displacements of the cation atoms from their symmetric positions in the centers of the lattice voids as well as niobium atom displacements from the centers of octahedra, which lowers the symmetry. The symmetry reduction induced by a possible ferroelectric distortion polarized along a given direction can be easily taken into account by eliminating the symmetry operations which do not keep this direction invariant. Thus determined subgroups are listed in Table 3.

**Table 3**. Space groups of ferroelastic-ferroelectric substructures for different polarization directions

| $a_s, b_s, c_s$ | Ferroelastic substructures | | Spontaneous polarization direction | | | | | |
|---|---|---|---|---|---|---|---|---|
| | | | $\mathbf{a}_s$ | | $\mathbf{b}_s$ | | $\mathbf{c}_s$ | |
| 1, 1, 2 | $D_{2h}^{16}$ | *Pbnm* (62) | $C_{2v}^2$ | *P*2$_1$*am* (26) | $C_{2v}^2$ | *Pb*2$_1$*m* (26) | $C_{2v}^9$ | *Pbn*2$_1$ (33) |
| $\sqrt{2}, \sqrt{2}, 2$ | $D_{2h}^{17}$ | *Ccmm* (63) | $C_{2v}^{14}$ | *C*2*mm* (38) | $C_{2v}^{14}$ | *Cm*2*m* (38) | $C_{2v}^{12}$ | *Ccm*2$_1$ (36) |
| $\sqrt{2}, \sqrt{2}, 2$ | $C_{4h}^5$ | *I*4/*m* (87) | $C_2^3$ | *I*112 (5) | $C_2^3$ | *I*112 (5) | $C_4^5$ | *I*4 (79) |
| $\sqrt{2}, \sqrt{2}, 2$ | $D_{2h}^{28}$ | *Ibmm* (74) | $C_{2v}^{20}$ | *I*2*mm* (44) | $C_{2v}^{22}$ | *Ic*2*m* (46) | $C_{2v}^{22}$ | *Ibm*2 (46) |
| 1, 2, 2 | $C_{2v}^{12}$ | *A*2$_1$*am* (36) | $C_{2v}^{12}$ | *A*2$_1$*am* (36) | $C_s^3$ | *A*11*m* (8) | $C_s^4$ | *A*1*a*1 (9) |
| 1, 2, 2 | $C_{2h}^3$ | *A*112/*m* (12) | $C_s^3$ | *A*11*m* (8) | $C_s^3$ | *A*11*m* (8) | $C_2^3$ | *A*112 (5) |
| $\sqrt{2}, \sqrt{2}, 2$ | $D_{2h}^{16}$ | *Pcmn* (62) | $C_{2v}^2$ | *P*2$_1$*ma* (26) | $C_{2v}^9$ | *Pc*2$_1$*n* (33) | $C_{2v}^2$ | *Pcm*2$_1$ (26) |
| 1, 2, 2 | $C_{2h}^2$ | *P*112$_1$/*m* (11) | $C_s^1$ | *P*11*m* (6) | $C_s^1$ | *P*11*m* (6) | $C_2^2$ | *P*112$_1$ (4) |
| 2, 2, 2 | $C_{2h}^1$ | *P*112/*m* (10) | $C_s^1$ | *P*11*m* (6) | $C_s^1$ | *P*11*m* (6) | $C_2^1$ | *P*112 (3) |
| $2\sqrt{2}, \sqrt{2}, 2$ | $D_{2h}^{17}$ | *Bbmm* (63) | $C_{2v}^{12}$ | *Bm*2*m* (35) | $C_{2v}^{12}$ | *Bb*2$_1$*m* (36) | $C_{2v}^{16}$ | *Bbm*2 (40) |
| $2\sqrt{2}, 2\sqrt{2}, 2$ | $C_{2h}^3$ | *B*112/*m* (12) | $C_s^3$ | *B*11*m* (8) | $C_s^3$ | *B*11*m* (8) | $C_2^3$ | *B*112 (5) |

*3.3. Overview of crystal structures of TTB-like niobates.*

Some of the structures listed in Table 3 can be found among crystal structures discovered experimentally for various TTB niobate compounds. Some of these compounds are listed in Table 4.

Table 4. List of the TTB-like crystalline niobates

| Compound | | Notation |
|---|---|---|
| $Ba_4Na_2$ | $Nb_{10}O_{30}$ | BNN |
| $(Ba_{0.67}Sr_{0.33})_5$ | $Nb_{10}O_{30}$ | BSN |
| $(Sr_{0.67}Ba_{0.33})_5$ | $Nb_{10}O_{30}$ | SBN |
| $Pb_4K_2$ | $Nb_{10}O_{30}$ | PKN |
| $Pb_2K_4Li_2$ | $Nb_{10}O_{30}$ | PKLN |
| $(Ca_{0.28}Ba_{0.72})_5$ | $Nb_{10}O_{30}$ | CBN |
| $(Ba_{0.67}Re_{0.33})_5$ | $(TiNb)_{10}O_{30}$ | BRTN |

Re = Bi, La, Nd, Sm and Gd

Crystal structures of these compounds are collected in Table 5.

Table 5. Crystal structures of TTB niobates according to experimental data

| | Compound | Space group | | $a_s, b_s, c_s$ | SP direction | order/disorder | Reference |
|---|---|---|---|---|---|---|---|
| | BNN | | | | | | [9] |
| 1 | BSN | P4bm | (100) | 1, 1, 1 | $c_s$ | o | [10] |
| | PKLN | | | | | | [11] |
| 2 | BNN | Pba2 | (32) | 1, 1, 1 | $c_s$ | o | [9] |
| | PKLN | | | | | | [15] |
| 3 | SBN | P4bm | (100) | 1, 1, 1 | $c_s$ | d | [12, 13] |
| 4 | BNN | Cmm2 | (35) | $\sqrt{2}, \sqrt{2}, 1$ | $c_s$ | d | [16] |
| 5 | BNN | Ccm21 | (36) | $\sqrt{2}, \sqrt{2}, 2$ | $c_s$ | o | [17] |
| | SBN | | | | | | [18] |
| 6 | SBN | Bmm2 | (38) | $\sqrt{2}, 2\sqrt{2}, 2$ | $c_s$ | O | [18] |
| | CBN | | | | | INC | [19] |
| 7 | PKN | Cm2m | (38) | $\sqrt{2}, \sqrt{2}, 1$ | $b_s$ | o | [20] |
| 8 | BNN | Bbm2 | (40) | $2\sqrt{2}, \sqrt{2}, 2$ | $c_s$ | o | [21] |
| 9 | SBN | Im2a | (46) | $2\sqrt{2}, 2\sqrt{2}, 2$ | $b_s$ | o | [22] |
| 10 | BRTN | Im2a | (46) | $\sqrt{2}, \sqrt{2}, 2$ | $c_s$ | o | [6] |
| 11 | SBN | P4bm | (100) | 1, 1, 2 | $c_s$ | INC | [14] |

Comparing Tables 5 and 3, one can draw the following conclusions:
1. The RUM-induced structures should be accompanied by a doubling of unit cell in c-direction. Hence, the structures 1-4 and 7 (numbers of lines in Table 5) cannot be directly related to any RUM-induced ferroelastic structure.
2. With the exception of structures shown in the first line of Table 3, unit cell parameters of the RUM-induced structures should differ from those of the parent tetragonal structure. Hence, the structures 1-3 and 11 cannot be directly related to any RUM-induced ferroelastic structure.
3. Structures 5, 8 and 9 can be truly considered as RUM-induced structures.
4. The structure $Bmm2$ is very close to the RUM-induced structure $Bbm2$. In fact, the latter was considered in Ref [18] as a possible alternative but was rejected (in our opinion erroneously).
5. The structure $Im2a$ is equivalent to the RUM-induced structure $Ic2m$ but corresponds to twice larger $a$ and $b$ parameters. This may be caused by a hidden commensurate structure modulation.

The RUM distortions are soft. They are numerous among the phonon states of a TTB-like lattice. Therefore, they can give rise to various structural fluctuations, static as well as dynamic. Such fluctuations may condensate in a regular periodic superstructure or may induce an INC modulation. Equally well, the structural fluctuations may lead to a state with a random spatial distribution having a static or a dynamic character. In such a case, they would result in unusually large and highly anisotropic atomic displacement parameters or even may appear in results of crystal structure resolution, as a spatial split of some atomic positions.

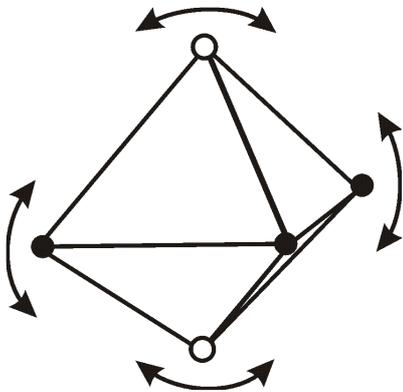

Figure 4. Tilting octahedron oscillation giving rise to RUMs. The apex and equatorial oxygen atoms are shown by open and bold circles respectively

Fig. 4 shows atomic displacements during the tilting oscillations of octahedra involved in RUMs. It is clear that the equatorial oxygen atoms (shown as bold circles) have maximal amplitudes in vertical direction (along *c*-axis) and the apex oxygen atoms (shown as open circles) have maximal amplitudes in horizontal direction (in *ab*-plane). One can find similar relations between atomic displacement parameters in the published TTB structures which do not involve the *c* parameter doubling [10, 11]. In all cases when the anisotropic mean square amplitudes were determined, the amplitudes of equatorial oxygen atoms were found markedly larger in *c*-direction and those of the apex oxygen atoms are larger in the perpendicular direction. This fact allows suggesting the presence of a hidden structural disorder, static or dynamic. The latter case implies large amplitude oscillations dictated by RUMs.

Structures with split oxygen positions merit a special discussion. The presence of such splittings suggests that the reported structure is in fact an average of two (or several) ordered structures corresponding to opposite signs of RUM distortions. A detailed analysis of such disordered structures leads to the determination of possible ordered constituents. The disordered *Cmm*2 structure (case of $Ba_2NaNb_5O_{15}$, BNN) proposed in ref [16] is an example. According to chosen combination of RUM's signs it can be represented as an average of two ordered *Ccm*21 structures or of two ordered *Bbm*2 structures.

In some studies, structures of the TTB crystals were determined as belonging to the *P4bm* space group. This is a ferroelectric structure with a polar axis parallel to (001). Such a structure may be thought to originate from paraelectric phase without any ferroelastic distortion. However, diffraction experimental data and subsequent structure refinements [10-12] show that in these structures the oxygen thermal amplitudes are considerably anisotropic, which is quite consistent with a mechanism of condensation of RUMs as discussed above. Moreover, the presence of an INC modulation was found in some SBN structures [6, 14, 23] and it is remarkable that the directions of this modulation were found to be (110) and $(1\bar{1}0)$. This is in full agreement with our suggestion of possible RUM-induced INC structure variations.

## 4. Summary and conclusion

Lattice dynamics calculations reveal the existence of rigid unit modes in TTB crystal lattices. These phonon modes can be represented as spatially modulated combinations of the elementary vibrations. These are concerted rotations of the octahedra localized in narrow layers perpendicular to (110) and $(1\bar{1}0)$ directions. This particularity results in the occurrence of two RUM phonon branches along R-S-R and X-S-L lines in the BZ of a TTB lattice.

The RUM-induced substructures, i.e. crystal structures resulting from an ideal tetragonal TTB-lattice owing to the condensation of different RUMs, are specified. The analysis of available experimental data for TTB-like niobate crystals showed that some of these structures do coincide with the RUM-induced structures. Other cases, with disordered atomic positions may be interpreted as due to the condensation of several RUMs and would thus correspond to structures averaged over several RUM-induced structures. This novel result allows suggesting that the proposed scheme of the RUM-induced structures will be useful for the unified classification of the TTB-related crystal structures.

**Acknowledgements**

The authors would like to thank V. Kazimirov and T. Smirnova for preparation of the figures. We thank also the Conseil Régional Languedoc Roussillon (France) for financing an invited professor position for one of us (M.S.). This study was partly supported by the Russian Foundation for Basic Research (project no. 12-03-01140-a).